\documentclass[aps,prb,showpacs,twocolumn,amsmath,amssymb,superscriptaddress]{revtex4-1}

\usepackage{graphicx}
\usepackage{multirow}
\usepackage{dcolumn}
\usepackage{bm}

\begin{document}

\title{Molecular Adsorption on Metal Surfaces with a van der Waals Density Functional}

\author{Guo Li}
\affiliation{International Center for Quantum Design of Functional Materials (ICQD)/Hefei National Laboratory for Physical Sciences at the Microscale (HFNL), University of Science and Technology of China, Hefei, Anhui, 230026, China}
\affiliation{Institute of Physics, Chinese Academy of Sciences, Beijing 100190, China}
\affiliation{Department of Physics and Astronomy, University of Tennesse, Knoxville, TN 37996, USA}

\author{Isaac Tamblyn}
\affiliation{Molecular Foundry, Lawrence Berkeley National Laboratory, 1 Cyclotron Road, Berkeley, CA 94720, USA}

\author{Valentino R. Cooper}
\affiliation{Materials Science and Technology Division, Oak Ridge National Laboratory, Oak Ridge, TN 37831, USA}

\author{Hong-Jun Gao}
\affiliation{Institute of Physics, Chinese Academy of Sciences, Beijing 100190, China}

\author{Jeffrey B. Neaton}
\affiliation{Molecular Foundry, Lawrence Berkeley National Laboratory, 1 Cyclotron Road, Berkeley, CA 94720, USA}

\begin{abstract}

The adsorption of
1,4-benzenediamine (BDA) on the Au(111) surface and azobenzene on the
Ag(111) surface is investigated using density functional theory (DFT)
with a non-local density functional (vdW-DF) and a semi-local
Perdew-Burke-Ernzerhof (PBE) functional. For BDA on Au(111), 
the inclusion of London dispersion interactions not only dramatically
enhances the molecule-substrate binding, resulting in adsorption energies
consistent with experimental results, but also significantly alters
the BDA binding geometry. For azobenzene on Ag(111), the vdW-DF produces superior
adsorption energies compared to those obtained with other dispersion
corrected DFT approaches. These results provide evidence for the
applicability of the vdW-DF method and serves as a practical benchmark
for the investigation of molecules adsorbed on noble metal surfaces.

\end{abstract}

\maketitle


Understanding the fundamental interactions that bind organic molecules to noble metal substrates is of crucial importance in molecular-scale electronics and self-assembly, where the competition between molecule-substrate and intermolecular interactions can lead to templated arrangements with specific spectroscopic and transport properties.\cite{Joachim05p8801,Bartels10p87} As the forces driving the formation of these organic-inorganic assemblies often include both local chemical bonding and non-specific long-range interactions, it is essential to have an accurate description of both contributions (e.g.~Ref~\onlinecite{Li11p}). While density functional theory (DFT) provides a many-particle framework that, in principle, incorporates both local and nonlocal interactions, common semi-local approximations to DFT neglect long-range attractive contributions to van der Waals interactions, so-called \lq\lq{}London dispersion forces\rq\rq{}.\cite{London37p8b, Stone96p} However, in recent years, progress has been made towards including London dispersion corrections within standard DFT. These approaches run the gamut from semi-empirical methods to the development of more accurate exchange-correlation functionals.\cite{Johnson09p1127, Cooper10p1417} Among these methods, a fully first-principles van der Waals density functional (vdW-DF)~\cite{Dion04p246401, Thonhauser07p125112, Roman-Perez09p0812.0244} has been developed to accurately include the effects of London dispersion forces. This method has been shown to be relatively accurate as well as computationally tractable and, as such, has been applied with success to a wide range of systems including the adsorption and wetting of various surfaces.\cite{Langreth09p084203a, Toyoda10p134703, Carrasco11p26101, Zhang11p236103}

In the present work, we perform DFT calculations, with and without long-range London dispersion corrections, to investigate the adsorption of 1,4-benzenediamine (BDA) and azobenzene on the (111) surfaces of Au and Ag. In the BDA/Au(111) system, we demonstrate that inclusion of dispersion forces via the recent vdW-DF results in enhancements to molecule-substrate binding, bringing predictions for adsorption energies into agreement with experimental results.~\cite{DellAngela10p2470a} Furthermore, use of the vdW-DF significantly alters the preferred orientation of the molecule relative to the Au(111) surface, resulting in a preference for a flat adsorption geometry over the tilted configuration previously obtained with a generalized gradient approximation (GGA).\cite{DellAngela10p2470a} Likewise, in the azobenzene/Ag(111) system, vdW-DF results in better adsorption energies than those obtained with other semi-empirical dispersion corrections. Compared with experiment\cite{Mercurio10p36102a} and complementing previous vdW-DF studies of flat molecules on Au(111) surfaces,\cite{Mura10p4759} our results provide evidence for the utility of vdW-DF for studies of molecule-metal binding. 


Our DFT calculations used both the vdW-DF\cite{Dion04p246401, Thonhauser07p125112} and the Perdew-Burke-Ernzerhof (PBE)\cite{Perdew96p3865} GGA to compare the effects of dispersion interactions on the adsorption of molecules to noble metal surfaces.
 All calculations of BDA molecules on Au(111) were performed using a 408 eV planewave cutoff and ultrasoft pseudopotentials\cite{Vanderbilt90p7892} as implemented in a modified version of the Quantum Espresso simulation package (QE ver. 4.2.1).\cite{Giannozzi09p395502} (The vdW-DF module was obtained from the SIESTA simulation package.\cite{Roman-Perez09p0812.0244}) Calculations of azobenzene molecules on Ag(111) were performed with the VASP (5.2.12) simulation package, employing a 500 eV plane-wave cut off and PAW potentials.\cite{Kresse99p1758,Kresse96p11169}  For both systems, a periodic 4-atom-layer slab with 20 \AA\ of vacuum and a 2$\times$2$\times$1 Monkhorst-Pack \textit{k}-point mesh were used; both are found to result in converged energetics and binding geometries. During relaxations, the bottom two layers are fixed and all other atoms were allowed to relax unconstrained until the forces on each atom were less than 3 meV/\AA. 

Our computed Au lattice constants are 4.14 \AA~(PBE), and 4.25 \AA~(vdW-DF), respectively. The overestimate of the lattice constant relative to experiment (4.08~\AA) within vdW-DF has been noted before\cite{Puzder06p164105,Vydrov08p14106,Romaner09p53010} and was attributed to excessive exchange (resulting in unphysically strong short range repulsion). We find similar agreement with experiment for the bulk Ag lattice constant yielding values of 4.16 \AA, and 4.26 \AA\ within the PBE and vdW-DF. To model the Au substrate, a 4-layer 4$\times$4 in-plane unit cell containing 64 Au atoms was used. Following Ref.~\onlinecite{Mercurio10p36102a}, we used a 4-layer Ag slab with a 3$\times$6 in-plane unit cell of 72 Ag atoms. 



BDA-Au junctions have been intensely investigated as a prototype for understanding charge transport at the molecular scale.\cite{Venkataraman06p458, Quek07p3477, Andrews08p1120, DellAngela10p2470a, Fatemi11p1988, Tsutsui09p10552, Strange11p115108, Kiguchi08p13349, Kiguchi10p22254, Mowbray08p111103} In a BDA-Au junction, amine groups preferentially bind to under-coordinated Au atoms, resulting in well-defined conductance.~\cite{Venkataraman06p458, Quek07p3477, Andrews08p1120,Tsutsui09p10552}  Temperature-dependent helium atom scattering experiments reported the binding energy (BE) of a related but distinct system, sparse BDA sub-monolayers adsorbed on flat Au(111) substrates, as roughly 1 eV,\cite{DellAngela10p2470a} stronger than the computed bond strength between an amine group and an under-coordinated Au atom (0.4-0.7 eV).\cite{Venkataraman06p458, Quek07p3477, Tsutsui09p10552} Moreover, flat Au(111) is expected to be chemically inert, and thus a primary contributor to the large adsorption energy is expected to be London dispersion interactions between the BDA molecule and the Au substrate. These interactions were not explicitly accounted for in previous calculations (see Ref.~\onlinecite{DellAngela10p2470a}). In what follows, we use calculations with the vdW-DF and the PBE functional to compare their performance with experiment, and to better understand the role of nonlocal, dispersion forces on BDA adsorption.

\begin{figure}[!h]
\centering
\includegraphics[width=3 in]{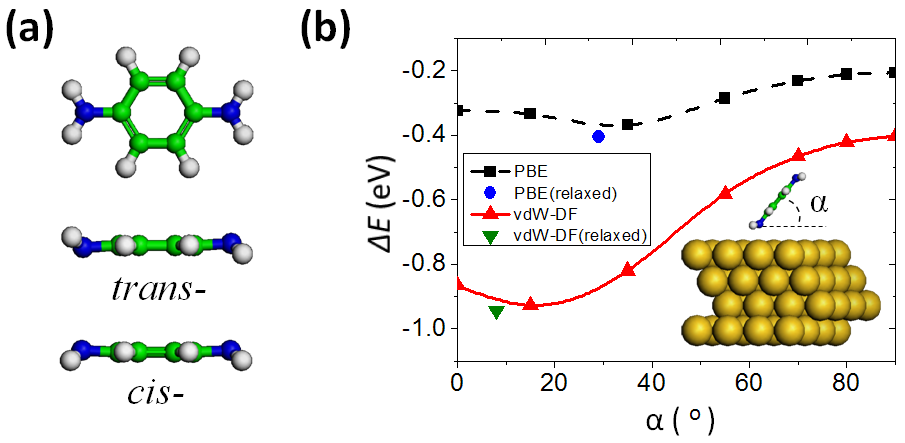}
\caption{(Color online). (a) The \textit{trans}- and \textit{cis}- conformation of BDA. (b) Dependence of adsorption energy on the tilt angle, $\alpha$, within PBE (black $\blacksquare$) and vdW-DF (red $ \blacktriangle$) for \textit{trans}-BDA on Au(111).}
\label{f:bda_angle_dependence}
\end{figure}

The amine groups at either end of the gas-phase BDA molecule adopt a pyramidal structure with two H atoms located on one side of the phenyl plane and an electron lone pair on the other. Consequently, BDA is stable in both the \textit{trans}- or in the \textit{cis}- structure (see Fig.~\ref{f:bda_angle_dependence}a).

Adsorption energies of a \textit{trans}-BDA on Au(111) as a function of tilt angle relative to the surface are shown in Fig.~\ref{f:bda_angle_dependence}b, where $\alpha$ denotes the angle between the phenyl plane of BDA and the Au surface. For each angle, each of the atoms within the BDA molecule is constrained to preserve the tilt angle with the surface. Using this approach, the PBE energetic minimum is found to be $\sim 35^\circ$ with a binding energy of 0.37 eV. Conversely, the vdW-DF calculations exhibited significant increases in the adsorption energies as well as a shift in the tilt angle to 15$^\circ$. Evidently the inclusion of nonlocal dispersion interactions results in the molecule being effectively pulled closer to the Au surface.

\begin{figure}[t]
\centering
\includegraphics[width=3 in]{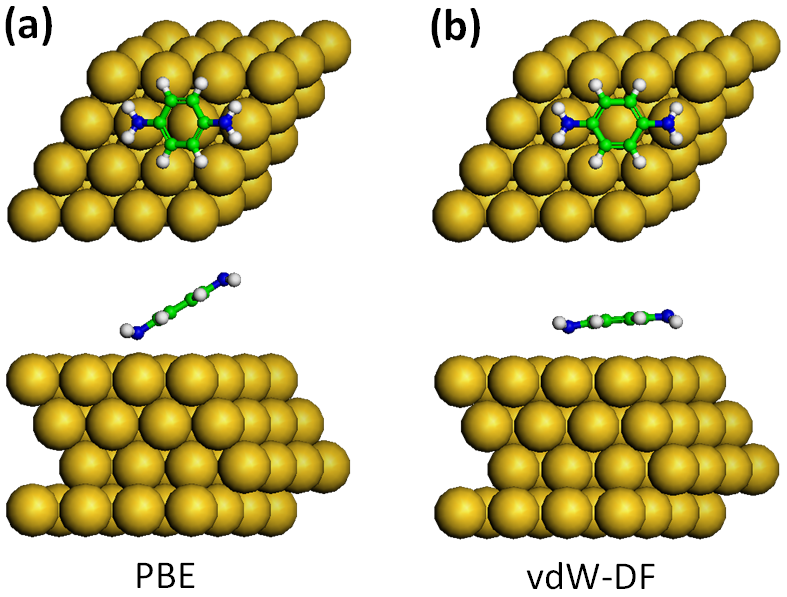}
\caption{(Color online). Equilibrium tilting configurations obtained within (a) PBE and (b) vdW-DF scheme for the \emph{trans} conformation of BDA on the Au(111) surface.}
\label{f:bda_tilt_relaxed}
\end{figure}

To obtain accurate equilibrium adsorption energies and optimized configurations, full structural relaxations are also performed. For PBE (see Fig.~\ref{f:bda_tilt_relaxed}(a)), we find that the adsorption energy is 0.41 eV with a N-Au surface atom distance of 2.55 \AA\ (at a tilt angle of 29$^\circ$). These results agree with those reported previously,\cite{DellAngela10p2470a} where the binding was attributed to a weak but non-negligible amine-Au bond. On the other hand, for vdW-DF (see Fig.~\ref{f:bda_tilt_relaxed}b) the adsorption energy is 0.94 eV, in good agreement with the experimental value of 1 eV. Here the N-Au distance is 3.12 \AA\, significantly larger than the PBE result. For the relaxed \emph{trans} molecule on the Au(111) surface, we find that the tilt angle is now 8$^\circ$ with the four amine H atoms located at the same height relative to the surface. This indicates that the phenyl plane tilt is attributable to the \emph{trans}-structure rather than the amine-Au interaction (see Fig.~\ref{f:bda_tilt_relaxed}b). 

\begin{figure}[t]
\centering
\includegraphics[width=3 in]{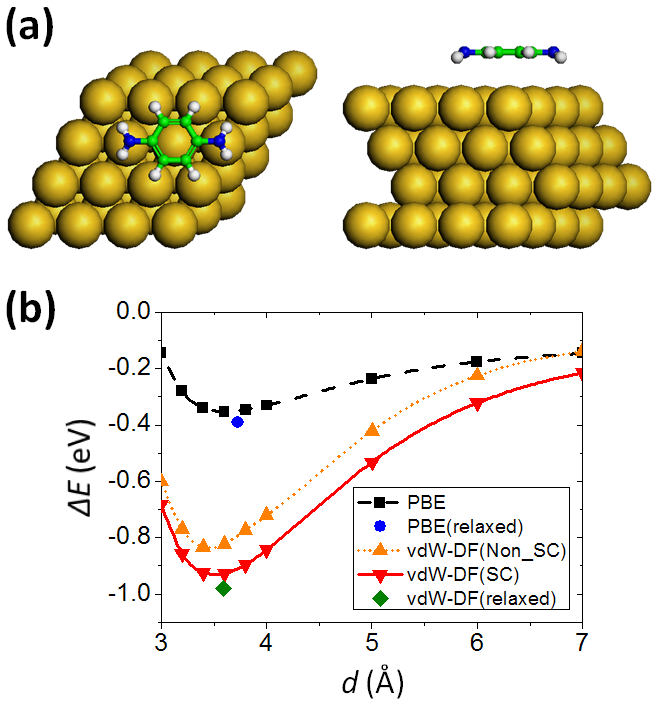}
\caption{(Color online).  Computed binding energy curves for the \emph{cis}-BDA on the Au(111) surface.  The vdW-DF (SC) results are computed using a fully self-consistent approach at the vdW-DF Au lattice constants. vdW-DF (non-SC) results are performed using the PBE charge density (structures and lattice constants are based on PBE), i.e. post-processing of vdW-DF.}
\label{f:bda_separation}
\end{figure}

BDA can also bind to the Au(111) surface (see Fig.~\ref{f:bda_separation}a) in the \emph{cis}-conformation.  In this case, we find that PBE and vdW-DF give similar lowest energy geometries, with the four N-bonded H atoms pointing towards the surface.  The dependence of adsorption energy on the BDA-Au(111) separation is illustrated in Fig.~\ref{f:bda_separation}b. PBE calculations find an optimum separation distance of 3.71 \AA\ with a binding energy of 0.39 eV, similar to the tilted structure discussed above. vdW-DF calculations predict an optimal separation of 3.57 \AA\ , and an adsorption energy of 0.98 eV, slightly stronger than that of the vdW-DF tilted \textit{trans}-configuration. The similar binding energies between the \emph{cis}- and \emph{trans}- conformations obtained using PBE, leading to meta-stable tilts, emphasize the fact that the molecule-surface interactions are dominated by dispersion forces, which stabilize flat adsorption geometries. Furthermore, the large enhancement in binding energies within vdW-DF (0.60 eV) further confirms the need for an accurate treatment of non-local interactions in investigations of metal-molecule interfaces.

Consideration of the adsorption of \textit{cis}-BDA at different sites on the Au(111) surface indicates that the Au(111) surface is energetically flat for BDA molecules, i.e. the adsorption energies for different sites are similar ($<$ 10 meV). Although the ordering of binding sites within vdW-DF differs from PBE, in both cases the differences between sites is very small, essentially on the order of the expected DFT error.  Similar results are expected for the \textit{trans}-configuration.



\begin{figure}[!b] 
\begin{center}
\includegraphics[width=0.5\textwidth,clip]{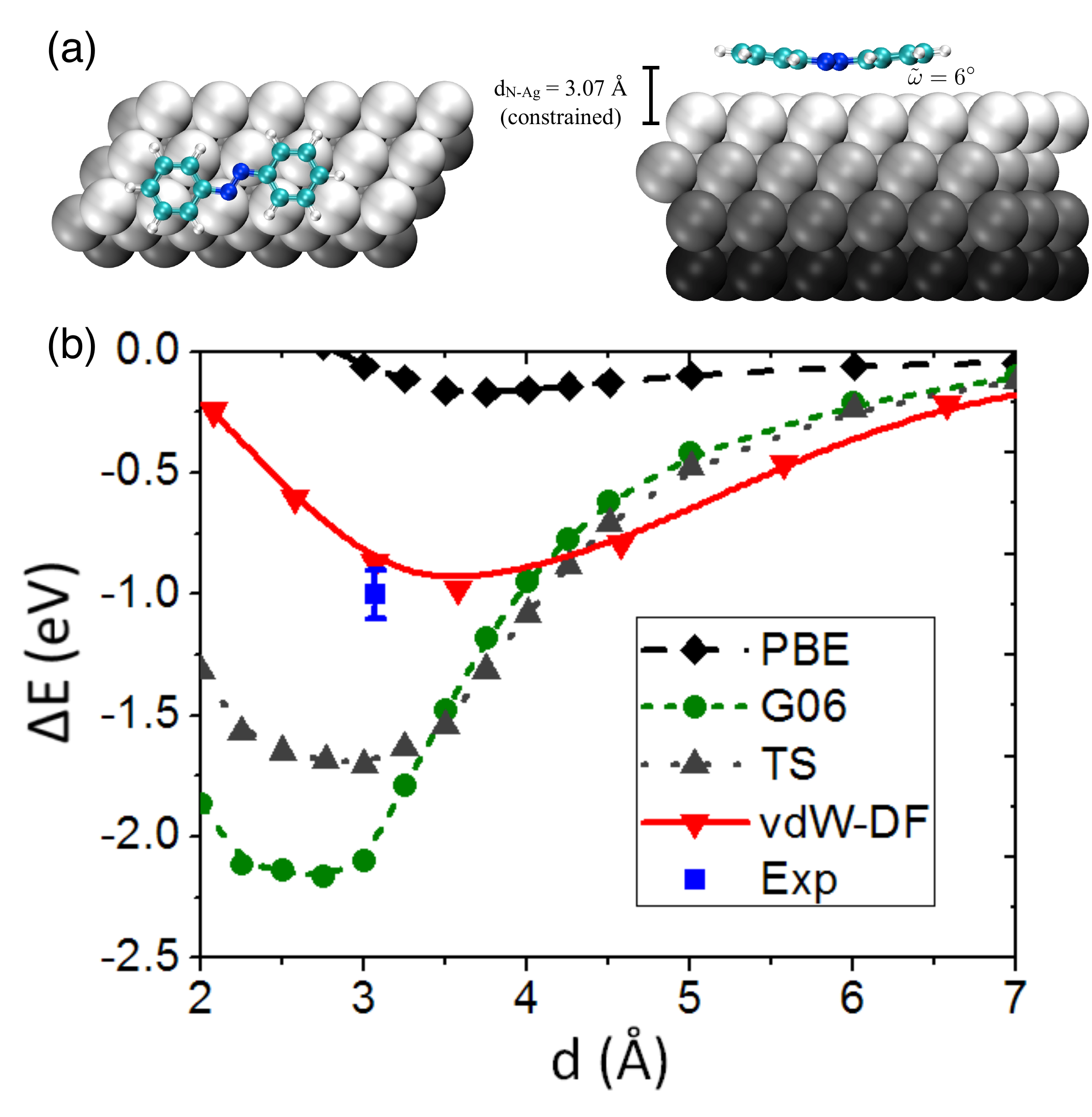}
\end{center}
\caption{\label{f:azobenzene_desorption} (Color online). Computed binding energy curves for azobenzene on Ag(111). Experimental and theoretical results from Ref.~\onlinecite{Mercurio10p36102a} are shown. vdW-DF (this work) produces a BE in excellent agreement with experiment. The adsorption height is however overestimated by 16\%, as discussed in the text.}
\end{figure}

To compare with our BDA-Au(111) calculations and further assess the efficacy of the vdW-DF, we also compute the equilibrium adsorption geometry and BE for a related molecule, azobenzene, on the Ag(111) surface. This system is chosen not only because its optical-structural properties have attracted significant attention,\cite{Comstock07p38301a} but also due to the recent availability of experimental measurements of its adsorption energetics. Mercurio \emph{et al.}~\cite{Mercurio10p36102a} reported both the binding geometry and adsorption energy of azobenzene on Ag(111) using a normal-incidence X-ray standing wave (NIXSW) approach. They find that upon adsorption the N-Ag distance is 3.07 $\pm$ 0.02 \AA \, and that the molecule lies flat relative to the surface ($\omega = -1 \pm 0.2$). Furthermore, they measured an adsorption energy of 1.0 eV. This value was obtained through an examination of the coverage dependence of thermally programmed desorption (TPD) experiments. Their work was also compared with several first principles approaches, including PBE-GGA \cite{Perdew96p3865} but also the dispersion-corrected methods of Grimme\cite{Grimme06p1787} and Tkatchenko-Scheffler\cite{Tkatchenko09p73005}. (These prior results are reproduced in Fig.~\ref{f:azobenzene_desorption} for comparison). 

Using vdW-DF, we compute an azobenzene-Ag(111) BE of 0.98 eV, in excellent agreement with experiment. This is in stark contrast with the small BE predicted by PBE (0.1 eV). The larger discrepancy between PBE and vdW-DF BE for azobenzene (compared with BDA) is consistent with its \emph{two} polarizable aromatic rings (rather than just one). Like PBE however, the vdW-DF predicted adsorption height is too large (by about 0.5 \AA). Although this is an improvement over PBE (a 26\% overestimate), the adsorption height is still an order of magnitude larger than typical errors associated with DFT bond lengths. This is consistent with vdW-DF\rq{}s typical overestimation of separation distances (a consequence of excessively repulsive exchange interactions~\cite{Klimes10p22201,Cooper10p161104,Lee10p81101}). Our computed value for $\omega$ (see Fig.~\ref{f:azobenzene_desorption}) is also consistent with the experimental measurement. As with BDA on Au, the site dependence of the BE seems to be small. Results are nearly identical when one of the N atoms is placed above an atop site (d$_{\textrm{N-Ag}}$ = 3.63 and BE=0.97 eV).

Since both the adsorption height and angle have been measured experimentally, we can also examine how well vdW-DF describes the interactions between the aromatic ring and the surface at fixed d$_{\textrm{N-Ag}}$. When we constrain d$_{\textrm{N-Ag}}$ to the experimental value and relax the system, the BE is reduced to 0.9 eV, and the aromatic rings tilt an angle of 6$^\circ$ relative to the surface. Thus, it appears that the interaction between the aromatic group is too repulsive within vdW-DF.

One possible source of the height overestimate may stem from the fact that within the vdW-DF functional, the equilibrium lattice constant for bulk Ag is 4.26~\AA ~(vdW-DF2 gives 4.32~\AA). This is larger than what PBE predicts (4.16 \AA), which is itself already too large compared to experiment (4.08 \AA). To test this hypothesis, we compute the vdW-DF equilibrium height and BE of azobenzene on a silver slab constructed with the smaller PBE lattice constant; we find that the adsorption height is 3.64~\AA, and the BE is 0.98 eV. Using a variant of vdW-DF which performs better at predicting the bulk lattice constant of Ag (optB86b+vdW-DF\cite{Klimes11p195131}, 4.11 \AA) results in poorer performance. Our computed BE and height with optB86b+vdW-DF are 1.54 eV and 2.85 \AA\ respectively. Similar results are obtained when applying the C09x exchange functional with vdW-DF (C09x+vdW-DF\cite{Cooper10p161104}) for BDA on the Au (111) surface.  Here, the C09x+vdW-DF gives an excellent Au lattice constant (4.09 \AA), but overbinds by the same order of magnitude as obtB86b-vdW-DF. This highlights the delicate balance between short and long range forces which must be achieved when developing transferable functionals capable of accurately describing heterogeneous interfaces.

In conclusion, we have used DFT calculations to investigate the influence of London dispersion interactions on the adsorption of 1,4-benzenediamine (BDA) on Au(111) and azobenzene on Ag(111). A non-local vdW density functional (vdW-DF) was used to determine adsorption energies and the corresponding molecular configurations. In the BDA/Au(111) system, we find that inclusion of vdW interactions produces adsorption energies consistent with experimental results, significantly enhancing molecule-substrate binding over PBE and stabilizing flat adsorption geometries. In the azobenzene/Ag(111) system, vdW-DF results in better binding energies compared to other dispersion corrected functionals, albeit with some overestimation of adsorption height ($\approx$16\%). This work provides evidence for the relevance of the vdW-DF approach for the structure and stability of metal-molecule binding.

\section*{Acknowledgement}

We thank T. Thonhauser and B. Kolb for use of their implementation of the SIESTA module in QE. This work was supported in part by the U.S. NSF (grant No. DMR-0906025), the NSF of China (grant Nos.  60921092 and 11034006), and by National ``973" programs of China. Calculations were carried out at NERSC. Work at the Molecular Foundry was supported by the Office of Science, Office of Basic Energy Sciences, of the U.S. Department of Energy under Contract No. DEAC02-05CH11231. I.T. acknowledges financial support from NSERC. V.R.C. was supported by the U.S. Department of Energy, Basic Energy Sciences, Materials Sciences and Engineering Division.


%

\end{document}